\documentclass[fleqn,twoside,10pt]{elsart}

\usepackage{graphicx}
\usepackage{units}
\usepackage{amsmath}
\input{pubboard/babarsym}
\newcommand{\rff}[1]{\ref{fig:#1}}
\newcommand{\rfe}[1]{Eq.~\ref{equ:#1}}
\newcommand{\eqp}{.}
\newcommand{\eqc}{,}
\newcommand{\pderiv}[2]{\frac{\partial #1}{\partial #2}}
\newcommand{\ie}{\emph{i.e.}}
\newcommand{\eg}{\emph{e.g.}}
\newcommand{\atantwo}{\ensuremath{\mathrm{atan2}}}

\begin{document}

\begin{frontmatter}

\title{Decay Chain Fitting with a Kalman Filter}
\author{Wouter D. Hulsbergen}
\ead{hulsberg@slac.stanford.edu}
\address{University of Maryland\\
  Mailing address: Stanford Linear Accelerator Center, 
  P.O. Box 20450, Stanford, CA 94309.}

\title{}


\begin{abstract}
  We present a method to perform a least squares fit of a decay chain
  involving multiple decay vertices.  Our technique allows for the
  simultaneous extraction of decay time, position and momentum
  parameters and their uncertainties and correlations for all
  particles in a decay chain.
\end{abstract}

\end{frontmatter}

\maketitle


\section{Introduction}

In high energy physics experiments decay reactions that proceed via
intermediate metastable states are usually reconstructed by following
a bottom-up approach. One starts by extracting the parameters of those
decay vertices from which the reconstructed final state particles
emerge and uses the intermediate `composite' particles for the
reconstruction of upstream decays. At each decay vertex the parameters
of the composite particle are determined with a least squares fit to
its daughter particles, subject to the constraint that those originate
from a common point. The disadvantage of this approach, which is
sometimes called `leaf-by-leaf' fitting, is that constraints that are
upstream of a decay vertex do not contribute to the knowledge of the
parameters of the vertex. An example of a decay for which this is
impractical is $\KS\to\piz\piz$.

In this paper we discuss the implementation of a least squares fit
that extracts all parameters in a decay chain simultaneously. We shall
call this fit, which we developed for data analysis in the \babar{}
experiment, a global decay chain fit. First, we propose a
parameterization of a decay chain in terms of vertex positions,
momenta and decay times. Subsequently, we argue that the Kalman filter
is a suitable technique to extract these parameters and the
corresponding covariance matrix from the external constraints, which
in the case of \babar{} are reconstructed charged particle
trajectories and neutral particle calorimeter clusters. Finally, we
present two examples and briefly summarize experience with the fit in
\babar{}.

The decay chain fits discussed here are hypothesis driven. The task of
finding the reconstructed tracks and clusters and associating those
with the final state particles in the decay tree is outside the scope
of this paper. In \babar{} physics analyses decay trees are built
layer-by-layer, usually by making all possible combinations of final
state particles and applying selections on the invariant mass and
vertex $\chi^2$. Vertex pattern recognition plays an insignificant
role because the low combinatorics does not warrant more complicated
algorithms and because the track parameter resolution is barely
sufficient to separate the decay vertices of the particles that are of
most interest to the experiment, namely $B$ and $D$ mesons.

\section{Parameterization of a decay tree \label{sec:parameterization}}

Figure~\rff{decaychain} shows a schematic picture of a decay tree. The
positions of the vertices in the decay tree, and the momenta of all
particles, constitute the degrees of freedom of the decay tree.  These
degrees of freedom, the internal constraints (such as momentum
conservation at each vertex) and the relation to the external
reconstruction objects, define the decay tree model.

\begin{figure}[htb]
  \includegraphics[width=\columnwidth]{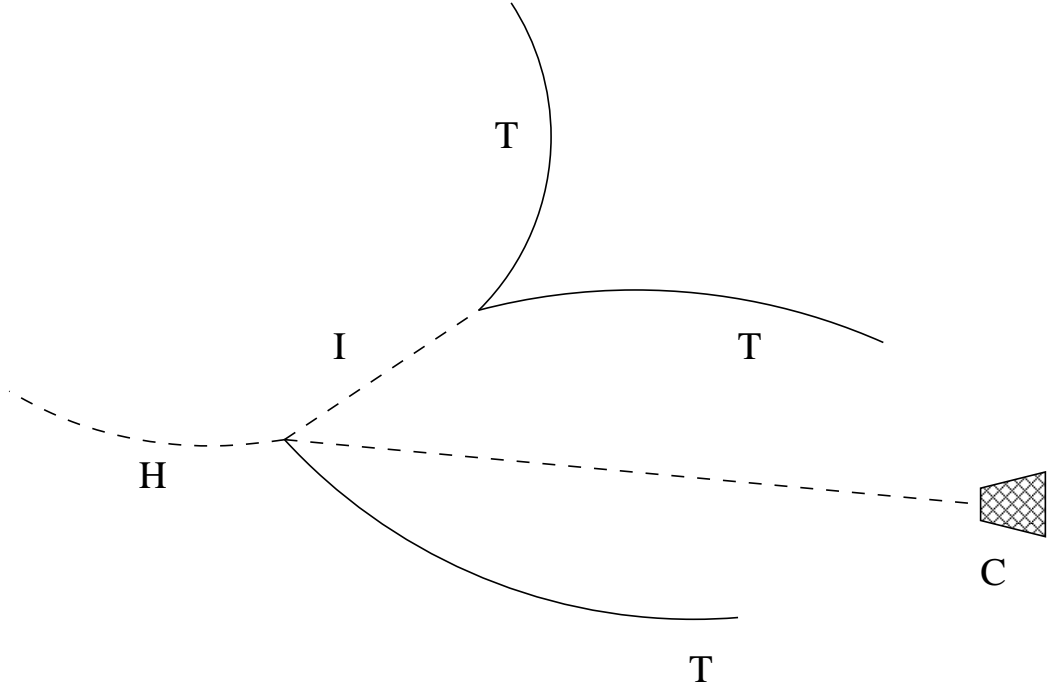}
  \caption{Schematic picture of a decay tree with three charged
    particles reconstructed as track segments (T), one photon
    reconstructed as a calorimeter cluster (C), and two composite
    particles (I for `internal' and H for `head').}
  \label{fig:decaychain}
\end{figure}

The choice of parameters in the decay tree model is not unique, but we
found the following parameterization suitable for use in \babar{}.
Each reconstructed or `final state' particle is represented by a
momentum vector $(p_x,p_y,p_z)$. The mass of a final state particle is
not a parameter in the fit, but assigned based on the particle
hypothesis in the decay tree.  Each intermediate particle in the decay
tree is modeled by a four momentum vector $(p_x,p_y,p_z,E)$ and a
decay vertex position $(x,y,z)$. If the composite particle is not at
the head of the decay tree, we also assign a parameter for its decay
time. We choose this parameter to be $\theta\equiv l/|\vec{p}|$, where
$l$ is the decay length.

If a composite particle has an expected decay length much smaller than
the vertex detector resolution, we call this particle a `resonance'.
A resonance does not have a decay time parameter and it shares the
decay vertex position with its mother, unless it is at the head of the
decay tree.  In the \babar{} reconstruction software particles with an
expected decay length \unit[$c\tau<1$]{$\mu$m}, such as \piz{},
\jpsi{} and \Dstar{}, are treated as resonances.\footnote{We use the
  term resonance for any particle with a short lifetime, regardless of
  whether the decay is through the strong or electroweak interaction.}

We distinguish two types of constraints in the decay tree. Two
\emph{internal} constraints are applied to remove redundant degrees of
freedom: the vertex constraint expresses the relation between the
decay vertex of a particle and the production vertex of its daughters;
the momentum conservation constraint ensures four-momentum
conservation at each vertex. The reconstructed final state particles
constitute the \emph{external} constraints. In this paper we consider
only 5-parameter track segments and calorimeter clusters with a
reconstructed position and energy.  Explicit expressions for the
constraints are given in section~\ref{sec:constraints}. To put those
in the proper context we introduce the fit procedure first.

\section{Fitting a decay tree \label{sec:chisqfit}}

\subsection{Measurement constraints and exact constraints}

To extract the optimal set of decay tree parameters from the
reconstruction information we use a least squares fit. Within this
framework constraints are expressed as $\chi^2$ contributions which
are a function of the parameters of the model, collectively denoted by
the vector $x$. The solution to the fit is the value of $x$ that
minimizes the total $\chi^2$. To define the $\chi^2$ contributions we
distinguish between \emph{exact} constraints and \emph{measurement}
constraints.  The latter are characterized by the fact that they have
an associated uncertainty, whereas the former do not. The internal
constraints of the decay tree model are exact constraints, whereas the
external constraints are measurement constraints.

For a measurement constraint $i$ the $\chi^2$ contribution takes the
form
\begin{equation}
  \chi^2_i \; = \; r_i^T(x) \;\;  V_i^{-1} \; r_i(x)
\end{equation}
where $r_i(x)$ is the constraint residual and $V_i$ is the constraint
variance. We use a matrix notation so that the $\chi^2$ contribution
is defined for constraints of any dimension. The residual can be
defined as
\begin{equation}\label{equ:residualdef}
  r_i(x) \; = \; m_i \: - \: h_i(x)
\end{equation}
where $m_i$ is the value of the measured quantity (for example the
parameters from a reconstructed track) and $h_i(x)$ is the measurement
model that expresses this quantity in terms of the parameters $x$.
With this definition $V_i$ is the variance in the measurement $m_i$.

Exact constraints can be imposed on the model by parameter
substitution. For example, the parameters that represent the momentum
of a mother particle can directly be replaced by a sum over the
daughter momenta. Although this simplifies the fitting problem by
reducing the number of parameters, it is not suitable for the decay
tree fit discussed here.  First, the substitution procedure makes it
more complicated to use recurrence in the implementation of the fitting
algorithm. Second, it complicates the calculation of those parameters
that have been removed --- the momentum of the mother ---, especially
where it concerns the associated uncertainty. 

Consequently, exact constraints are not applied by substitution but by
means of Lagrange multipliers. If the exact constraint in terms of
$x$ is written as
\begin{equation}
  g_i(x) \; = \; 0 \eqc
  \label{equ:exactconstraint}
\end{equation}
the $\chi^2$ contribution of the constraint takes the form
\begin{equation}\label{equ:chisqexact}
  \chi^2_i \; = \; 2 \: \lambda_i^{T} \: g_i(x)
\end{equation}
where $\lambda_i$ is the Lagrange multiplier. The latter is added to
the list of parameters $x$ for which we minimize the $\chi^2$.

\subsection{The standard least squares fit}

The requirement that $\chi^2 = \sum_i \chi^2_i$ be minimal defines a
set of equations $\partial\chi^2/\partial x=0$. A solution can be
obtained with a Newton-Raphson method. (See for
example~\cite{bock:1998}.) Given an initial estimate $x^{(0)}$, one
obtains a new estimate
\begin{equation}
   x^{(1)} \; = \; x^{(0)} - \left(\pderiv{^2\chi^2}{x^2}\right)^{-1}
            \pderiv{\chi^2}{x} \eqp
\end{equation}
For constraints that are not linear in $x$ this expression can be
applied iteratively until a certain convergence criterion is met.
Using error propagation one can derive that the variance in $x$ is
given by
\begin{equation}
  C(x) \; = \; 2 \left(\pderiv{^2\chi^2}{x^2}\right)^{-1} \eqp
\end{equation}

We shall call the procedure described above, in which the $\chi^2$
contributions of all constraints are minimized simultaneously, the
\emph{standard} fit.\footnote{This method for $\chi^2$ minimization is
  often called a `global' fit, but we reserve the term global in
  this paper for use in `global decay chain fit', a decay chain fit
  that determine all parameters and correlations in a decay tree
  simultaneously, rather than leaf-by-leaf.}  The expressions for the
standard fit show that the minimization procedure requires the
inversion of matrices with the dimension of the parameter vector $x$.
Complicated decay chains can easily require several tens of
parameters, which leads to large computational costs.

\subsection{The progressive least squares fit}

The Kalman filter or \emph{progressive} fit~\cite{Kalman:1960} is a
$\chi^2$ minimization method that is less computation intensive than
the standard fit. In the reconstruction of data from particle physics
experiments the Kalman filter is mainly applied in track fitting,
where its virtue is not only speed, but also the possibility to
easily include multiple scattering effects as so called `process
noise'~\cite{Fruhwirth:1987fi}. The Kalman filter has been proposed
for vertex fitting by several other
authors~\cite{Fruhwirth:1987fi,Luchsinger:1992ia}.  However, a general
treatment of exact constraints has not been considered before, which
justifies a re-derivation of the Kalman filter for the purpose of
decay chain fitting. To connect with previous work we follow a
notation close to that in~\cite{Fruhwirth:1987fi}.

Consider a measurement constraint $k$. Assume that the $\chi^2$
consisting of all contributions of constraints $\{0,\ldots,k-1\}$ has
already been minimized with respect to the model parameters, leading
to a `prediction' $x_{k-1}$ with variance $C_{k-1}$. To calculate how
$x$ changes when the constraint $k$ is included in the minimization, 
we define a $\chi^2$ contribution
\begin{eqnarray}
  \label{equ:chisquare}
  \chi^2_k & \; = \; &   
  ( x - x_{k-1})^{T} \: {C_{k-1}}^{-1} \: ( x - x_{k-1})
    \nonumber\\
    & & \; + \;  ( h_k(x) - m_k )^{T} \: {V_k}^{-1} \: ( h_k(x) - m_k ) \eqp
\end{eqnarray}
The least squares solution for $x$ is obtained from the requirement that
$\partial\chi^2/\partial x \equiv 0$, \ie
\begin{equation}\label{equ:minChi2condition}
      {C_{k-1}}^{-1} ( x - x_{k-1}) \; + \; 
      H_k^{T} \: V_k^{-1} \: ( h_k(x) - m_k ) \; = \; 0
\end{equation}
where $H_k \equiv \partial h/\partial x|_{x_{k-1}}$ is the derivative
or projection matrix. We call the solution to this equation
the \emph{updated} parameter vector $x_k$.

The authors of reference~\cite{Fruhwirth:1987fi} discusses two
approaches to calculate $x_k$, which are called the `gain matrix
formalism' and the `weighted means formalism'. The latter still
requires inversion of matrices with the dimension of $x$ and is not
suitable for our purposes. In the gain formalism one rewrites
\rfe{minChi2condition} as
\begin{equation}
  \left(  {C_{k-1}}^{-1} +     
    H_k^{T} {V_k}^{-1} H_k \right) ( x - x_{k-1}) \; = \;
  H_k^{T} {V_k}^{-1} r_k^{k-1}
\end{equation}
where we defined the so-called residual of the prediction
\begin{equation}
  r_k^{k-1} =  m_k - h_k(x_{k-1})
\end{equation}
and where we assumed that the measurement model is linear, \ie{}
\begin{equation}
  h_k(x) \; = \; h_k(x_{k-1}) \: + \: H_k \: ( x - x_{k-1}) \eqp
\end{equation}
The treatment of more general $h$ will be discussed below. Solving for
$x$, we obtain
\begin{equation}\label{equ:filteredpar}
   x_k \; = \; x_{k-1} \: + \: K_k \: r_k^{k-1}
\end{equation}
where the gain matrix $K_k$ is defined as
\begin{equation}\label{equ:gainmatrix}
    K_k \; = \; 
    \left(  {C_{k-1}}^{-1} +  H_k^{T} {V_k}^{-1} H_k \right)^{-1} H_k^{T} {V_k}^{-1} \eqp
\end{equation}
The latter can be rewritten as
\begin{equation}\label{equ:simplegainmatrix}
    K_k \; = \; C_{k-1} H_k^{T} \left(R_k^{k-1}\right)^{-1}
\end{equation}
where 
\begin{equation}
    R_k^{k-1} \; = \; V_k + H_k C_{k-1} H_k^{T}
\end{equation}
is the uncertainty in the predicted residual. The fact that
\rfe{simplegainmatrix} contains the inverse of a matrix with the
dimension of the measurement $m_k$, rather than with the dimension of
the parameter vector $x$, is the reason that the progressive fit is in
general faster than the standard fit.

The updated covariance matrix $C_k$ can be obtained by error
propagation from \rfe{filteredpar}, which gives
\begin{equation}\label{equ:filteredCov}
  C_k \; = \; 
  (1 - K_k H_k) \: C_{k-1} \: (1 - K_k H_k)^{T} 
  \; + \; K_k V_k {K_k}^{T}\eqp
\end{equation}
This expression is computation intensive because the first term on the
right hand side corresponds to a product of three square matrices with
the dimension of the parameter vector. It can be simplified to
\begin{equation}\label{equ:filteredCovSimple}
  C_k \; = \; \left( 1 \:- \: K_k H_k \right) C_{k-1}
\end{equation}
which is much faster, but known to be sensitive to finite machine
digit accuracy, in particular if $V_k$ is small compared to $H_k C_{k-1}
H_k^{T}$~\cite{Gelb:1974}. This can be understood by evaluating the
effect of a small perturbation in the gain matrix $K_k \rightarrow K_k
+ \delta K$. Substituting this in~\rfe{filteredCov} yields $\delta C_k
= \delta K R_k \delta K^T$, whereas~\rfe{filteredCovSimple} gives
$\delta C_k = -\delta K H_k C_k$. As a result the second expression
can lead to a covariance matrix with a negative determinant, an effect
that we indeed observed in fits with many parameters. We have found
that by rewriting \rfe{filteredCov} as
\begin{equation}\label{equ:filteredCovAlt}
  C_k \; = \; C_{k-1} \; - \; K_k \left( 2 H_k C_{k-1}  \: - \: R_k^{k-1} K_k^{T} \right)\eqc
\end{equation}
the computational stability is preserved at the cost of a moderate
increase in computation time with respect to \rfe{filteredCovSimple}.

Finally, the $\chi^2$ contribution of the constraint $m_k$ is given by
the $\chi^2$ contribution of the prediction, \ie{}
\begin{equation}
  \chi^2_k \; = \; (r_k^{k-1})^{T} (R_k^{k-1})^{-1} r_k^{k-1}\eqp
\end{equation}
The $\chi^2$ contribution of a particular constraint depends on the
order in which the constraints are applied. However, if all
constraints are linear in $x$, the sum of the $\chi^2$ contributions is
independent of that order.

\subsection{Non-linear constraints \label{sec:nonlinearfilter}}

If the measurement model is not linear in $x$, $x_k$ can be extracted
with an iterative procedure. Expanding $h(x)$ around a point $x^{(i)}$
(initially given by $x_{k-1}$),
\begin{equation}
  h_k(x) \; = \; h_k(x^{(i)}) \: + \: H_k^{(i)} (x - x^{(i)})
\end{equation}
the estimate for $x_k$ becomes
\begin{equation}
  x_k^{(i+1)} \; = \; x_{k-1} \: + \: K_k^{(i)} \: r_k^{k-1(i)}
\end{equation}
where the residual $r_k^{k-1(i)}$ is given by
\begin{equation}
  r_k^{k-1(i)} \; = \;  m_k - h_k(x^{(i)}) \: - \: H_k^{(i)} (x_{k-1} -
  x^{(i)}) \eqp
\end{equation}
Choosing $x_k^{(i+1)}$ as the new expansion point, one obtains an
improved estimate of $x_k$ by iteration, subject to a certain
convergence criterion. A suitable observable to test the convergence
is the $\chi^2$ contribution
\begin{equation}
  \chi_k^{2\:(i)} \; = \; \left(r_k^{k-1(i)}\right)^T \left( R_k^{k-1 (i)}\right)^{-1}r_k^{k-1(i)}\eqp
\end{equation}
The time consuming calculation of the covariance matrix
(\rfe{filteredCov}) can be performed after convergence is obtained.

\subsection{Exact constraints \label{sec:exactconstraints}}

The $\chi^2$ contribution for an exact constraint $g(x)=0$ was
introduced in \rfe{chisqexact}.  Minimization of the $\chi^2$ under
the exact constraint is performed by solving the set of equations
$\partial\chi^2/\partial x = 0$ and $\partial\chi^2/\partial\lambda =
0$ for $x$ and $\lambda$ simultaneously. To derive the expressions for
the progressive fit we define a $\chi^2$
\begin{equation}
  \chi_k^2 \; = \; 
  ( x - x_{k-1})^{T} { C_{k-1} }^{-1} ( x - x_{k-1} ) \; + \; 2 \lambda_k^{T} \: g_k(x) \eqp
\end{equation}
Linearizing the constraint equation around $x_{k-1}$
\begin{equation}
  g_k(x) \; = \; g_k(x_{k-1}) \: + \: G_k \: (x - x_{k-1}) \: + \: \cdots 
\end{equation}
where $G = \partial g/\partial x$, we obtain for the linearized set
of first derivatives
\begin{equation}
  \begin{array}{lll}
    0 & \; = \; & {C_{k-1}}^{-1} ( x - x_{k-1}) \; + \; G_k^{T} \lambda_k \\
    0 & \; = \; & g_k(x_{k-1}) \; + \; G_k \: (x - x_{k-1})
  \end{array}
\end{equation}
Multiplying the first equation by $G C_{k-1}$ and subtracting the second
equation yields
\begin{equation}
  (G_k C_{k-1} G_k^{T}) \: \lambda_k \; = \; g_k(x_{k-1}) \eqp
\end{equation}
Since for non-trivial constraint equations the matrix on the left side
is invertible, this equation leads to a solution for $\lambda_k$.
Eliminating $\lambda_k$ we obtain for the updated parameter vector
\begin{equation}\label{equ:filteredparexact}
  x_k \; = \; x_{k-1} \; - \; K_k \: g_k(x_{k-1})
\end{equation}
where we defined the gain matrix for exact constraints by
\begin{equation}
  K_k \; = \; C_{k-1} G_k^{T} \left( G_k C_{k-1} G_k^{T} \right)^{-1} \eqp
\end{equation}
Using error propagation we obtain for the updated covariance matrix
\begin{equation}
  C_k \; = \;  (1-K_k G_k) \: C_{k-1} \: (1-K_k G_k)^T 
\end{equation}
and for the $\chi^2$ contribution
\begin{equation}
  \chi_k^2 \; = \; g_k(x_{k-1})^{T}  \left( G_k C_{k-1} G_k^{T} \right)^{-1} g_k(x_{k-1}) \eqp
\end{equation}
A comparison of the equations for the exact constraint with
those for the measurement constraint shows that with the substitutions
\begin{equation}
  \begin{array}{cclll}
    h(x_{k-1}) - m_k  & \longrightarrow &  g_k(x_{k-1}) \\
    V_k + H_k C_{k-1} H_k^{T} & \longrightarrow & G_k C_{k-1} G_k^{T} \\
  \end{array}
\end{equation}
the two procedures are identical. This non-trivial result confirms the
intuitive notion that an exact constraint is effectively the same as a
measurement with infinite precision.

The minus sign in equation~\rfe{filteredparexact} with respect to its
counterpart~\rfe{filteredpar} is the result of our decision to choose
a conventional notation. It is customary to express the Kalman filter
equations in terms of the derivate of $h(x)$ to $x$ rather than the
derivate of the residual $r(x)$ to $x$.  The minus sign reflects the
fact that these derivatives differ by a sign when the residual is
defined as in~\rfe{residualdef}.  For the implementation of the fit
we have defined the residual with opposite sign, such that the same
expressions can be used for exact constraints and measurement
constraints.

It is interesting that the progressive fit deals more effectively with
exact constraints than the standard fit does. In the latter the
Lagrange multipliers are added explicitly as parameters to the fit,
increasing the number of parameters and therefore computational costs.
The treatment of an exact constraint in the standard fit is therefore
relatively expensive.  In the progressive fit the Lagrange multipliers
can be eliminated and exact constraints are less expensive than
measurement constraints.

\section{Explicit expressions for constraints in the decay tree 
  \label{sec:constraints}}

In this section we provide explicit expressions for the internal
and external constraints introduced in section~\ref{sec:parameterization}.

\subsection{The internal constraints}

The internal constraints reduce the set of decay tree parameters to a
set that is not overcomplete. In the absence of a magnetic field or
for neutral particles, the momentum vector $\vec{p}$ is constant,
leading to a time-evolution of the position $\vec{x}(t) = \vec{x}_0 \:
+ \: \vec{p} \: t/\gamma m$, where $t$ is the time in the lab frame, $m$
the rest mass and $x_0$ the position at $t=0$.  For each composite
particle $i$ in the decay tree the vertex constraint expresses the
relation between its decay vertex $\vec{x}_i$, its momentum
$\vec{p}_i$ and the decay vertex of its mother $\vec{x}_M$,
\begin{equation}
  \vec{x}_M - \vec{x}_i + \theta_i \vec{p}_i \; = \; 0
\end{equation}
where $\theta \equiv l/|\vec{p}| = t /\gamma m$ is the decay time
parameter. The momentum constraint for particle $i$ can be expressed
as
\begin{equation}
  \vec{p}_i - \sum_j \vec{p}_j \; = \; 0 \qquad \text{and} 
  \qquad
  E_i - \sum_j E_j \; = \; 0 
  \eqc
\end{equation}
where the sum runs over the momenta of all the daughters.

For charged particles in a magnetic field the expressions for the
internal constraints are more complicated. In most practical
applications one can neglect the effect of the magnetic field for the
particles inside the decay tree, because the bending in a typical
lifetime is very small. This is for example the case for $B$ and $D$
mesons in the $1.5$-T magnetic field of the \babar{} detector. It is
not true for some charged baryons such as $\Xi^\pm$ and $\Sigma^\pm$.
The full expressions for the internal constraints in a constant
magnetic field have been implemented for \babar{} but are outside the
scope of this paper.

\subsection{The reconstructed track constraint}

The magnetic field in the \babar{} spectrometer is approximately
homogeneous and aligned with the $z$-axis of the detector.
Reconstructed charged particle trajectories are represented by a 5
parameter helix $m^T \equiv (d_0, \phi_0, \omega, z_0, \tan\lambda )$ and a
corresponding covariance matrix. In terms of these parameters, the
position $\vec{x}$ and momentum $\vec{p}$ of the particle along the
helix trajectory are given by
\begin{equation}
  \begin{array}{lll}
    x   & = &  r \sin(\phi) - (r+d_0)\sin(\phi_0) \\
    y   & = & -r \cos(\phi) + (r+d_0)\cos(\phi_0) \\
    z   & = & z_0 + l \tan(\lambda) \\
    p_x & = & p_t \cos(\phi) \\
    p_y & = & p_t \sin(\phi) \\
    p_z & = & p_t \tan(\lambda)
  \end{array}
\end{equation}
where $l$ is the flight length in the transverse plane, measured from
the point of the helix closest to the $z$-axis, $\phi = \phi_0 +
\omega l$, $r = 1/\omega$ and $p_t = q a/\omega$ with $q$ the charge
of the particle in units of the positron charge and $a[J/m] = -e[C]
B_z[T]$ a constant. The measurement model $h(x)$ used in the decay
tree fit follows from the inverse transformation and can be written
as
\begin{equation}
  h \; \equiv \;
  \left(
    \begin{array}{c}
      d_0         \\        
      \phi_0      \\      
      \omega      \\         
      z_0         \\     
      \tan\lambda \\ 
    \end{array}
  \right)
  \; = \; 
  \left(
    \begin{array}{c}
      ( p_{t0} - p_t ) / a q \\
      \atantwo(p_{y0}, p_{x0}) \\
      a q/p_t \\
      z - l p_z/p_t \\
      p_z/p_t \\
    \end{array}
  \right)
\end{equation}
with $p_t = \sqrt{p_x^2 +p_y^2}$, $p_{x0} = p_x + aqy$, $ p_{y0} = p_y
- aqx$, $p_{t0} = \sqrt{p_{x0}^2 +p_{y0}^2}$, $\phi
=\atantwo(p_y,p_x)$ and $l = ( \phi -\phi_0) p_t/qa$. The derivatives
can be concisely written as
\begin{equation}
  H^T \; \equiv \;
  \left( \begin{array}{c}
      \pderiv{h^T}{x} \\
      \pderiv{h^T}{y} \\
      \pderiv{h^T}{z} \\
      \pderiv{h^T}{p_x} \\
      \pderiv{h^T}{p_y} \\
      \pderiv{h^T}{p_z} \\
    \end{array} \right)
  \; = \; 
  \left( 
    \begin{array}{ccccc}
      \frac{-p_{y0}}{p_{t0}} &
      \frac{-aq p_{x0}}{p_{t0}^2} &
      0 &
      \frac{-p_z p_{x0}}{p_{t0}^2} &
      0 \\
      \frac{ p_{x0}}{p_{t0}} &
      \frac{-aq p_{y0}}{p_{t0}^2} &
      0 &
      \frac{-p_z p_{y0}}{p_{t0}^2} &
      0 \\
      0 & 0 & 0 & 1 & 0 \\
      \frac{1}{aq}\left(\frac{ p_{x0}}{p_{t0}} - \frac{ p_{x}}{p_{t}}\right) &
      \frac{-p_{y0}}{p_{t0}^2} & 
      \frac{-aq p_x}{p_t^3} & 
      \frac{-p_z}{aq} \left( \frac{p_{y0}}{p_{t0}^2} - \frac{p_y}{p_t^2} \right) &
      \frac{-p_z p_x}{p_t^3} \\
      \frac{1}{aq}\left(\frac{p_{y0}}{p_{t0}} - \frac{p_{y}}{p_{t}}\right) &
      \frac{ p_{x0}}{p_{t0}^2} &
      \frac{-aq p_y}{p_t^3} &
      \frac{ p_z}{aq} \left( \frac{p_{x0}}{p_{t0}^2} - \frac{p_x}{p_t^2} \right) &
      \frac{-p_z p_y}{p_t^3} \\
      0 & 0 & 0 & -\frac{l}{p_t} & \frac{1}{p_t} \\
    \end{array}
  \right).
\end{equation}

\subsection{The reconstructed cluster constraint}

In \babar{} reconstructed calorimeter clusters are represented by a
measured position and energy $m^T =
(\vec{x}_\text{clus},E_\text{clus})$ and a corresponding covariance
matrix $V_\text{clus}$.  Given a mother decay vertex $\vec{x}$ and a
momentum vector $\vec{p}$, the measurement model is defined as
\begin{equation}
  h =  \left( \begin{array}{c}
      x + \theta p_x \\
      y + \theta p_y \\
      z + \theta p_z\\
      \sqrt{p_x^2+p_y^2+p_z^2} \\
    \end{array} \right)
\end{equation}
where $\theta$ takes the role of the photon `decay length'. In
principle, $\theta$ can be added to the photon parameter list and
extracted from the fit. Since this parameter is not very interesting,
it is preferable to eliminate it. This can be done by reducing the set
of four constraint equations $r(x) \equiv m - h(x) = 0$ to three by
redefining the residual, for example
\begin{equation}
  r'(x) \; \equiv \;
  \left( \begin{array}{c}
      (x_\text{clus}-x) p_y - (y_\text{clus}-y) p_x \\
      (x_\text{clus}-x) p_z - (z_\text{clus}-z) p_x \\
      E_\text{clus} - \sqrt{p_x^2+p_y^2+p_z^2} \\
    \end{array} \right)
  \eqp
\end{equation}
The choice of $r'$ is not unique: This particular choice is not
suitable if $p_x$ is zero, because the two position constraints would no
longer be independent.  However, for every value of $\vec{p}$ a set of
independent constraints can be constructed. The variance of the new
constraint is simply given by
\begin{equation}
  V' \; = \; P \: V_\text{clus} \: P^{T}
\end{equation}
where $P=\partial r'/\partial m$ is the derivate of the constraint
to the measurement.

\subsection{Other constraints}

The constraints discussed above constitute the minimal set of
constraints necessary for fitting a decay tree with final state
particles that are reconstructed either as charged tracks or as
calorimeter clusters. We have considered and implemented several other
constraints. Sometimes, position and momentum parameters can be
improved by constraining the mass of composites in the decay tree to
the known particle mass. Knowledge about the interaction point can be
used to constrain the production vertex of the head of the decay tree.
Information about the beam momenta can be used to constrain the
four-vector of the head of the decay tree, \eg{} in
$\epem\to\Y4S\to\Bz\Bzb$ decays. In addition to charged tracks and
photons, we have used reconstructed $K^{0}_{L}$ particles which are
detected in the \babar{} calorimeter, but for which the deposited
energy is not useful as an estimate of the magnitude of the momentum.
Finally, we have found that missing particles can be included in the
decay tree, provided the tree is not kinematically under-constrained.
The expressions for these constraints are easy to derive so we will
not include them in this paper.

\subsection{Ordering constraints}

A disadvantage of the progressive fit with respect to the standard fit
is that the final result of the fit can be sensitive to the order in
which constraints are applied. In fits with `process noise' such as
track fits, there is a natural ordering.  However, in fits without
process noise, the ordering of the constraints can be chosen freely.

The order in which constraints that are linear in $x$ are applied is
irrelevant, because the covariance matrix contains all essential
information on the constraint derivatives. If, for example, a
constraint $k$ is processed such that $g_k(x_k)=0$, then $g_k(x_n)=0$
for any $n>k$. This is not the case for non-linear constraints, since
the covariance matrix does not contain information on the higher
derivatives of the constraint equation.  Consequently, the order in
which constraints are applied becomes important: The most non-linear
constraints should be applied last.

One can consider applying correlated non-linear constraints
simultaneously, rather than consecutively. For example, the vertex
constraint can be applied separately in the three space coordinates
$x$, $y$ and $z$.  However, the dependence on the decay time
parameter $\theta$ correlates these constraints.  Since they are also
slightly non-linear, it is preferable to treat them as a single
three-dimensional constraint rather than three separate
one-dimensional constraints.  Ultimately, all constraints can be
applied simultaneously, which effectively reduces the progressive fit
to a standard fit. The performance advantage of the progressive fit
then practically disappears.\footnote{The progressive fit still has
  the advantage that exact constraints do not lead to extra (Lagrange
  multiplier) parameters.}

Based on these considerations we have chosen the following approach to
ordering and combining constraint. The external constraints
(reconstructed tracks and cluster, constraints to the interaction
point etc.) are treated first. Subsequently, all four-momentum
conservation constraints are applied, starting at the end of the decay
tree.  Finally, at each vertex the geometric constraints and an
eventual mass are combined.  These combined constraints are applied
consecutively starting at the end of the decay tree.  

\subsection{Fit initialization and convergence}

The progressive fit requires initialization of both the parameters
$x_0$ and the covariance matrix $C_0$. Vertex positions are
initialized with the average interaction point or with the point of
closest approach of (a subset of) reconstructed track segments in the
decay tree. The momenta of particles reconstructed as a track segments
are initialized by evaluating the track parameters at the point of
closest approach to the initial vertex positions. The momenta of
photons are initialized by using the initial vertex positions as their
origin. Particle momenta inside the decay tree are initialized by
adding the initial four vectors of their daughters. Finally, the decay
time parameters are initialized from the initial vertex positions and
momenta or from the expected decay time.

The covariance matrix must be initialized with uncertainties that are
large enough that the initial parameters have a negligible weight in
the final result of the fit, yet small enough that the measurement
errors $V_k$ in equation are not numerically negligible with respect
to $H_k C_k H_k^T$. We have chosen for a diagonal matrix with diagonal
elements that are roughly a factor $1000$ larger than the square of
the typical resolution for the corresponding parameter.

Even with the iteration of non-linear constraints described in
section~\ref{sec:nonlinearfilter} the decay chain fit does not
converge in a single processing of all constraints.  Therefore, we
repeat the fit procedure until the total $\chi^2$ is stable. At each
step $x_0$ is initialized with the result of the previous step, while
$C_0$ is reset to its original value. Fits typically converge in three
iterations.

\section{Experience in \babar{} \label{sec:conclusions}}

We have implemented the decay chain fit described in the previous
sections in the \babar{} analysis framework. The fit has been
extensively tested and is used in a variety of physics analyses. It is
a useful alternative for leaf-by-leaf vertex fits, in particular for
the reconstruction of decay chains with vertices with insufficient
downstream constraints, such as $\Xi^0\to\Lambda^0\piz$ and
$\KS\to\piz\piz$, consistent treatment of mass constraints at several
levels in a decay tree, direct extraction of the $\Bz\Bzb$ decay time
difference for analysis of time-dependent \CP{} violation in
$\Y4S\to\Bz\Bzb$ events and kinematic fits with missing particles.  We
briefly discuss two examples.

\begin{figure}[htb]
  \centerline{
    \includegraphics[width=0.7\textwidth]{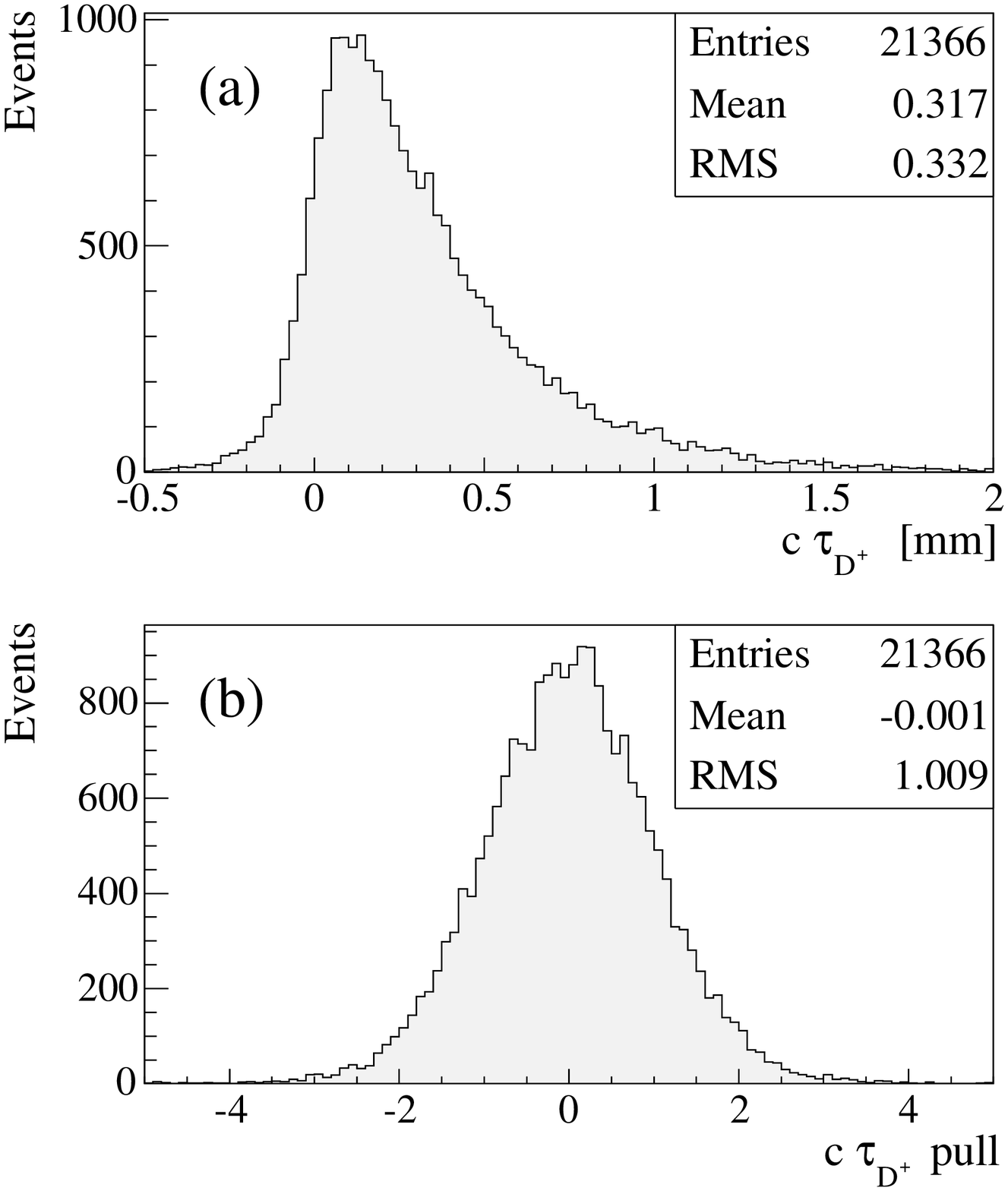}}
  \caption{Reconstructed lifetime (a) and lifetime pull (b) for 
    \Dpm{} extracted from a global decay chain fit to simulated
    $\Bzb\to\Dp\pim$ ($\Dp\to\Km\pip\pip$) decays.}
  \label{fig:btodpi}
\end{figure}

Figure~\ref{fig:btodpi}a shows the $c\tau$ distribution of $\Dp$
candidates in a \babar{} simulation of $\Bzb\to\Dp\pim$ decays with
$\Dp\to\Km\pip\pip$. Only candidates that are matched to the Monte
Carlo truth are shown. The average reconstructed lifetime is close to
the average lifetime \unit[$c\tau_{D^+}=0.311$]{mm} with which the
events were generated. Figure~\ref{fig:btodpi}b shows the $c\tau$ pull
distribution, which has an RMS consistent with one.

As a second example, we study the reconstruction of $\KS\to\piz\piz$
candidates in simulated $\Bz\to\jpsi\KS$ decays with
$\piz\to\gamma\gamma$ and $\jpsi\to\mumu$. The $\KS\to\piz\piz$ decay
cannot be reconstructed with \babar{}'s traditional leaf-by-leaf fit.
A global decay tree fit can be performed, provided that the origin of
the \KS{} is known and that mass constraints are applied to the \piz{}
candidates. In this particular topology the $\jpsi$ vertex provides
the origin and the $\Bz\to\jpsi\KS$ fit is thus sufficiently
constrained.

\begin{figure}[htb]
  \centerline{
    \includegraphics[width=0.7\textwidth]{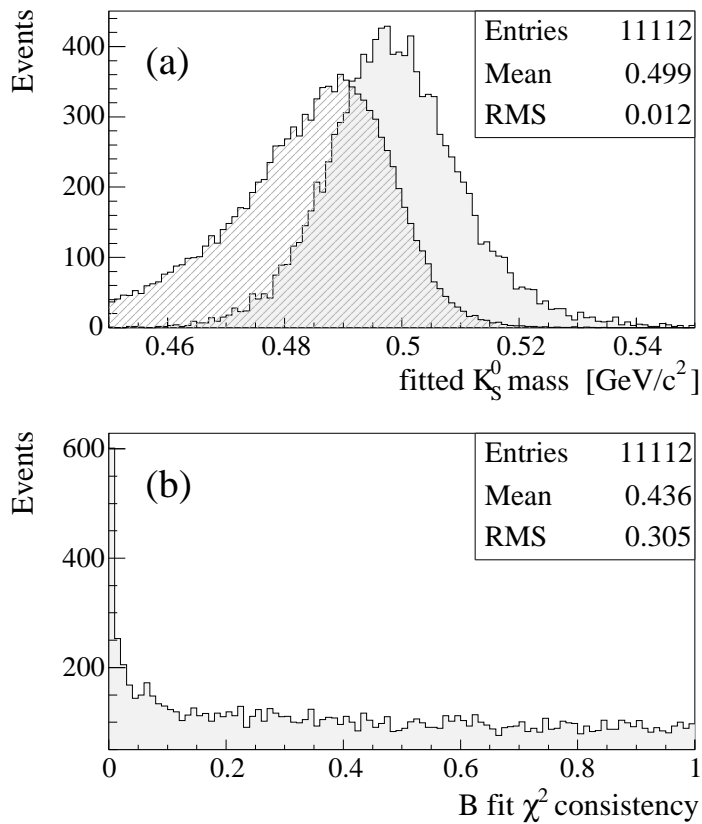}}
  \caption{$\KS\to\piz\piz$ invariant mass (a) extracted from a 
    global decay chain fit to simulated $\Bz\to\jpsi\KS$ decays and
    the $\chi^2$ consistency of the fit (b). The hashed distribution
    in figure (a) is the $\piz\piz$ invariant mass before the
    geometric fit is performed.}
  \label{fig:kstopizpiz}
\end{figure}

Figure~\ref{fig:kstopizpiz}a shows the invariant mass distribution of
the $\KS\to\piz\piz$ candidates extracted from a fit to the
$\Bz\to\jpsi\KS$ decay tree. The central value and resolution are
significantly better than for the `raw' $\piz\piz$ invariant mass
distribution that was obtained by assuming that the $\piz$ decays
originate directly from the $\jpsi\to\mumu$ vertex.
Figure~\ref{fig:kstopizpiz}b shows the $\chi^2$ consistency of the
$\Bz\to\jpsi\KS$ decay tree fit. (The $\chi^2$ has two degrees of
freedom.) It is not entirely consistent with a flat distribution,
partially because the energy of the photons originating from the
\piz{} candidates is not always fully reconstructed in the \babar{}
calorimeter.

The performance of the fit has been compared to a traditional
leaf-by-leaf vertex fit that has been used in \babar{} for several
years. A direct comparison of computational performance is not trivial
since that depends on implementation choices in addition to the
algorithmic complexity of the problem. The complexity of the global
decay chain fit grows roughly with the square of the number of
vertices in the decay chain, whereas the leaf-by-leaf fits (which do
not keep track of correlations) behave more linear.  Despite this
fact, computation time use has not been a limitation in practical
applications of the fit.

\section{Acknowledgment}

The author would like to thank Dr.~D.N.~Brown, Dr.~W.T.~Ford and
Dr.~A.~Jawahery for their help in preparing this document. The work
described here was supported by the U.S. Department of Energy under
grant number DEFG02-96ER41015.



\end{document}